\documentclass[aps,pre,twocolumn,amsmath,amssymb]{revtex4}

\usepackage{graphicx}
\usepackage{enumitem}
\usepackage{multirow,dcolumn}
\usepackage{newtxtext,newtxmath}

\begin{document}

\title{Epidemic extinction in networks: \\
Insights from the 12,110 smallest graphs}

\author{Petter Holme}
\affiliation{Institute of Innovative Research, Tokyo Institute of Technology, Nagatsuta-cho 4259, Midori-ku, Yokohama, Kanagawa, 226-8503, Japan}
\author{Liubov Tupikina}
\affiliation{Laboratoire de Physique de la Mati\`ere Condens\'ee (UMR 7643), CNRS -- Ecole Polytechnique, 91128 Palaiseau, France}

\begin{abstract}
We investigate the expected time to extinction in the susceptible-infectious-susceptible (SIS) model of disease spreading. Rather than using stochastic simulations, or asymptotic calculations in network models, we solve the extinction time exactly for all connected graphs with three to eight vertices. This approach enables us to discover descriptive relations that would be impossible with stochastic simulations. It also helps us discovering graphs and configurations of S and I with anomalous behaviors with respect to disease spreading. We find that for large transmission rates the extinction time is independent of the configurations, just dependent on the graph. In this limit, the number of vertices and edges determine the extinction time very accurately (deviations primarily coming from the fluctuations in degrees). We find that the rankings of configurations with respect to extinction times at low and high transmission rates are correlated at low prevalences and negatively correlated for high prevalences. The most important structural factor determining this ranking is the degrees of the infectious vertices.
\end{abstract}

\maketitle

\section{Introduction}

The susceptible-infectious-susceptible (SIS) model is the canonical model of infectious diseases that leave people re-susceptible to the disease upon recovery. As other compartmental models of infectious diseases~\cite{hethcote,andersonmay}, it consists of two main components. First, a local description of how the disease spreads between pairs of people, and dies. A susceptible individual in contact with an infectious individual becomes infectious with a rate $\beta$; infectious persons become re-susceptible with a rate $\nu$. Second, every epidemic model also describes how people come in contact with each other. Traditionally, one have assumed a fully-connected, or well-mixed, scenario---that anyone can meet anyone else with the same chance at all times. Lately, it has become popular to assume the population is connected into a network and everyone connected by an edge have equal probability of meeting one another, while pairs with no edge will never meet.

Research on the SIS model typically focuses on one of three questions. First, in a finite population how long time does it take for the outbreak to die out~\cite{o2,nasell,schwartz,doering,fagani}? Second, in an infinite population, there will be a threshold value of $\beta$ (given $\nu$) below which the outbreak inevitably dies out and above which it can live forever. This line of research investigates how the network structure---the probability distribution of degree (the number of neighbors), the number of triangles, etc.---affects the threshold~\cite{ps_rmp}. In almost all cases (Ref.~\cite{shadows} being an exception) authors have explored the large-size limit by stochastic simulations or approximative calculations. Other  questions include the ranking of important vertices with respect to the outbreak~\cite{qu_etal} and the chains of events that are most likely to lead to extinction~\cite{hindes}.

In this work, we will investigate a mix of the questions above. Namely how the network structure and the position of infectious vertices affect extinction in small connected graphs. To our knowledge, this is the first study to investigate the time to extinction from different configurations (of who is susceptible and who is infectious). Scanning small graphs, however, has occasionally been used in network science~\cite{masuda,holme3f,heetae}. Rather than addressing these questions with stochastic simulations, we calculate the exact expression for the expected time to extinction as a function of $\beta$ (we set $\nu=1$ without loss of generality). This approach is computationally expensive, so we restrict ourselves to connected graphs of eight vertices or less. On the other hand, we are not restricted by graph models but can go through every distinct (non-isomorphic) such graph; 12,110 in total.

Our non-stochastic computational approach makes it possible to discover exact relations among small graphs, such as: what the smallest graph is such that the ranking of configurations' extinction times is independent of $\beta$. We can also discover scaling relations, whose validity on one hand is only verified for small graphs, on the other hand cover the large-$\beta$ regime that is inaccessible for stochastic simulations. In general, the study of small graphs could be seen as a complement to the (more common) large-scale studies. Of course, the large-scale limit is a limiting case just like small networks are. Once can argue that some types of networks are so small that the lack of self-averaging of the large networks, makes this approach just wrong. Animal trade networks~\cite{Bajardirsif20120289}, for example, could be represented as a graph where the nodes are farms (technically speaking metapopulations). These are often small by design, to restrict outbreaks.

In the rest of the paper, we will describe our approach, in parallel present one example and introduce the general theory. Then we will go through the numerical findings, first in the limit of large $\beta$ and finally study how the ranking of configurations of susceptible and infectious nodes depend on $\beta$.

\section{Preliminaries}

\subsection{The SIS model}

Assume a graph $G=(V,E)$ with $N$ vertices labeled from $0$ to $N-1$. Let $\phi_i$ be a binary state variable ($\phi_i\in\{0,1\}$). We will interpret $\phi_i=0$ as vertex $i$ being susceptible while $\phi_i=1$ means that $i$ is infectious. In the common formulation of the SIS model~\cite{daleygani}, the probability of a susceptible vertex $i$ being infected by an infectious neighbor $j$ is $\beta$ per time unit, independent of when $j$ was infectious. Likewise, the recovery of $j$ is time independent, leading to an exponential distribution of the duration of infections. Without loss of generality, we can take the recovery rate to unity. This lack of memory (i.e.\ Markov property) means that we can encode the current situation of the outbreak into a number
\begin{equation}\label{eq:s_def}
 s=\sum_{i=0}^{N-1}\phi_i 2^i.
\end{equation}
Giving $s\in[0,2^N-1]$. Just like reading the string of S and I as 0 and 1 and interpreting it as a binary number. We will refer to $s$ as a \textit{configuration} of vertex states.

Next we will proceed to set up the equations for the expected time to extinction from a certain configuration. The derivation closely follows the derivation of the master equations (or Kolmogorov equations) giving the probability of the system being in a certain configuration~\cite{daleygani,kiss}. We thus effectively treat the SIS dynamics as a random walk in the space of configurations $s$, where $s=0$ is an absorbing configuration~\cite{masuda:rev}.

Now let $I(s)$ be all configurations reachable from $s$ by an infection event and $S(s)$ the set of configurations reachable from $s$ by a recovery. Let $\omega_s=|S(s)|$ be the number of infectious vertices, a.k.a.\ the \textit{prevalence}. Let $m_{st}$ be the number of edges between an infectious vertex in configuration $s$ and the vertex that is susceptible in $s$ and infectious in $t$ (in our encoding of configurations, this vertex is $\log_2 (t-s)$). Because of the exponential distribution of the durations in the susceptible and infectious states, the rates of events are additive. The total event rate $z_s(\beta)$ is 
\begin{equation}
z_s(\beta) = \beta\sum_{t\in I(s)}m_{st}+\omega_s,
\end{equation}
where $m_{st}$ is the number of infection events that would turn $s$ into $t$. This gives the expected duration of configuration $s$ as $1/z_s(\beta)$. The probability that the next configuration becomes $t$ via a infection event is $\beta \, m_{st}/z_s(\beta)$, while the probability of the next configuration $t$ reachable through a recovery event is $1/z_s(\beta)$, see Fig.~\ref{fig:ex}.

\subsection{Expected time to extinction}
\label{sec:exptime}
Consider a graph $G$. Let $x_s$ denote the expected time to extinction from configuration $s$. We can write down self-consistency equations for $x$ by noting it is the expected life time of the configuration $s$, $T_s=1/z_s(\beta)$, plus the expected extinction times of the configurations reachable from $s$ times their transition probabilities. Symbolically:
\begin{equation}
 x_s = T_s + \sum_t x_t \times \mathrm{Prob} (s\rightarrow t) , s\in[1, 2^N-1] .
\end{equation}
By the elementary laws of probability and the probabilities given in the previous section, this equation becomes
\begin{subequations}\label{eq:self}
\begin{align}
z_s(\beta) x_s &= 1 + \beta\sum_{t\in I(s)} m_{st} x_t + \sum_{t\in S(s)} x_t, ~ s>0\\
x_0&=0 .
\end{align}
\end{subequations}
From the above equation we can write the equation in the matrix form
\begin{equation}\label{eq:self2}
\mathbf{U}(\beta){\bf x}  + {\bf 1} = 0
\end{equation}
where ${\bf 1}=(1,\dots,1)^T$, ${\bf x} =(x_0,\dots x_{2^N-1}) $, and $\mathbf{U}(\beta)$ is a \emph{polynomial matrix}~\cite{Gantmacher} (since some of the its elements depend on $\beta$ parameter) defined by:
\begin{equation}\label{eq:y_mat}
U_{st}(\beta) = \left\{\begin{array}{ll} 1 & \mbox{if $s-t=2^i$, $i\in V$}\\
\beta m_{st} & \mbox{if $s\neq 0$ and $t-s=2^i$, $i\in V$}\\
-z_s(\beta) & \mbox{if $s=t$}\\
0 & \mbox{otherwise}
\end{array} \right.
\end{equation}
where we use the property that $s-t=2^i$, $i\in V$, if and only if the only difference between $s$ and $t$ is that vertex with number $i$ is infectious in $s$ and susceptible in $t$.

Extending Eq.~\eqref{eq:self} for all configurations $s$ generates a linear system of equations with as many equations as unknowns. We can thus solve it (we use Gaussian elimination in favor of more elaborate methods~\cite{mcclellan}) to get the expectation value of the extinction times from any initial configuration $s$.

For the example in Fig.~\ref{fig:ex}, Eq.~\eqref{eq:self} becomes:
\begin{subequations}\label{eq_subsyst}
\begin{align} 
(\beta+1)x_1 & = & 1 + \beta x_3 \\
(2\beta+1)x_2 & = & 1 + \beta x_3 + \beta x_6\\
(\beta+2)x_3 & = & 1 + x_1 + x_2 + \beta x_7\\
(\beta+1)x_4 & = & 1 + \beta x_6 \\
(2\beta+2)x_5 & = & 1 + x_1 + x_4 + 2\beta x_7 \\
(\beta+2)x_6 & = & 1 + x_2 + x_4 + \beta x_7\\
3x_7 & = & 1 + x_3 + x_6 + x_5 ,
\end{align}
\end{subequations}
where we have omitted the trivial $x_0=0$.
One can reduce this equation system further by grouping automorphically equivalent configurations (i.e.\ configurations that can be mapped to one another by a relabeling of the vertices)~\cite{Simon2011}. In the example of Fig.~\eqref{fig:ex}, configurations 1 and 4, and 3 and 6, form two automorphic equivalence classes. This reduces the equation system to:
\begin{subequations}\label{eq:redsys}
\begin{align}
(\beta+1)x_{1,4} & = & 1 + \beta x_{3,6} \\
(2\beta+1)x_2 & = & 1 + 2\beta x_{3,6}\\
(\beta+2)x_{3,6} & = & 1 + x_{1,4} + x_2 + \beta x_7\\
(2\beta+2)x_5 & = & 1 + 2x_{1,4} + 2\beta x_7 \\
3x_7 & = & 1 + 2x_{3,6} + x_5 ,
\end{align}
\end{subequations}
which, furthermore, gives a reduced version of ${\bf U}$ that we call ${\bf Y}$
\begin{equation}
\label{eq:ymat}
{\bf Y}(\beta) =
\begin{bmatrix}
-\beta-1 & 0 & \beta & 0 & 0 \\
0 & -2\beta-1 & 2\beta & 0 & 0 \\
1 & 1 & -\beta - 2 & 0 & \beta \\
2 & 0 & 0 & -2\beta-2 & 2\beta\\
0 & 0 & 2 & 1 & -3
\end{bmatrix} .
\end{equation}
Eq.~\eqref{eq:self2} holds with ${\bf U}$ replaced by ${\bf Y}$. Some properties of the matrix ${\bf Y}$ that hold for any network include:
\begin{enumerate}
\item \label{item:below} Below the diagonal, all elements are $\beta$ independent.
\item \label{item:above} Above the diagonal, the elements are integers times $\beta$.
\item \label{item:row} At each row, except the last (corresponding to the all-infectious configuration) there are two $\beta$-dependent elements. The constant coefficients of these terms sum to zero.
\item \label{item:diag} The diagonal is such that rows sum to zero, except the rows representing states that can reach $s=0$ by one recovery event, then the row sum is $-1$.
\end{enumerate}
Moreover, we note that the number of automorphic equivalence classes $n$ defines the rank of the reduced matrix.

Our example system in Eq.~\eqref{eq:redsys} has the solution:
\begin{subequations}\label{eq:ex_res}
\begin{align}
x_{1,4} & = & \frac{4\beta^4+16\beta^3+35\beta^2+34\beta+12}{16\beta^2+28\beta+12} \\
x_2 & = & \frac{4\beta^4+18\beta^3+42\beta^2+40\beta+12}{16\beta^2+28\beta+12} \\
x_{3,6} & = & \frac{4\beta^4+20\beta^3+51\beta^2+53\beta+18}{16\beta^2+28\beta+12}\\
x_5 & = & \frac{4\beta^4+20\beta^3+53\beta^2+52\beta+18}{16\beta^2+28\beta+12} \\
x_7 & = & \frac{4\beta^4+20\beta^3+57\beta^2+62\beta+22}{16\beta^2+28\beta+12} .
\end{align}
\end{subequations}
The expressions for $x_2$ and $x_{3,6}$ can be further simplified, but for comparison, we keep the same denominator.

\subsection{Algebraic calculations}

Solving Eq.~\eqref{eq:self2} is computationally complex. The major bottleneck is the polynomial algebra (to be precise---calculating the greatest common divisor needed to reduce the fractions of polynomials to their canonical form). The code was implemented in C with the FLINT library~\cite{flint} for polynomial algebra. To group automorphically equivalent configurations, it also relies on the subgraph-isomorphism algorithm VF2~\cite{vf2} as implemented in the igraph C library~\cite{igraph}. Finding subgraph isomorphisms---although a classical, computationally hard problem---is in practice relatively quick and this enables us to discover and exploit all symmetries rather than \textit{a priori} focusing on symmetrical graphs (cf.\ Ref.~\cite{kiss}).

Our code is available at \url{github.com/pholme/sis_exact/}.

\subsection{Small distinct graphs}

We systematically evaluate small distinct (non-isomorphic) connected graphs of sizes up to $8$ vertices: $3\leq N\leq 8$. There are two such graphs with $N=3$, six with $N=4$, 20 with $N=5$, 112 with $N=6$, 853 with $N=7$ and 11,117 with $N=8$, in total, $12,110$ graphs for $3\leq N\leq 8$ vertices. To generate these, we use the program Geng~\cite{McKay201494}. They can also be downloaded and viewed at \url{http://www.graphclasses.org/smallgraphs.html}.

\begin{figure}
\includegraphics[width=\columnwidth]{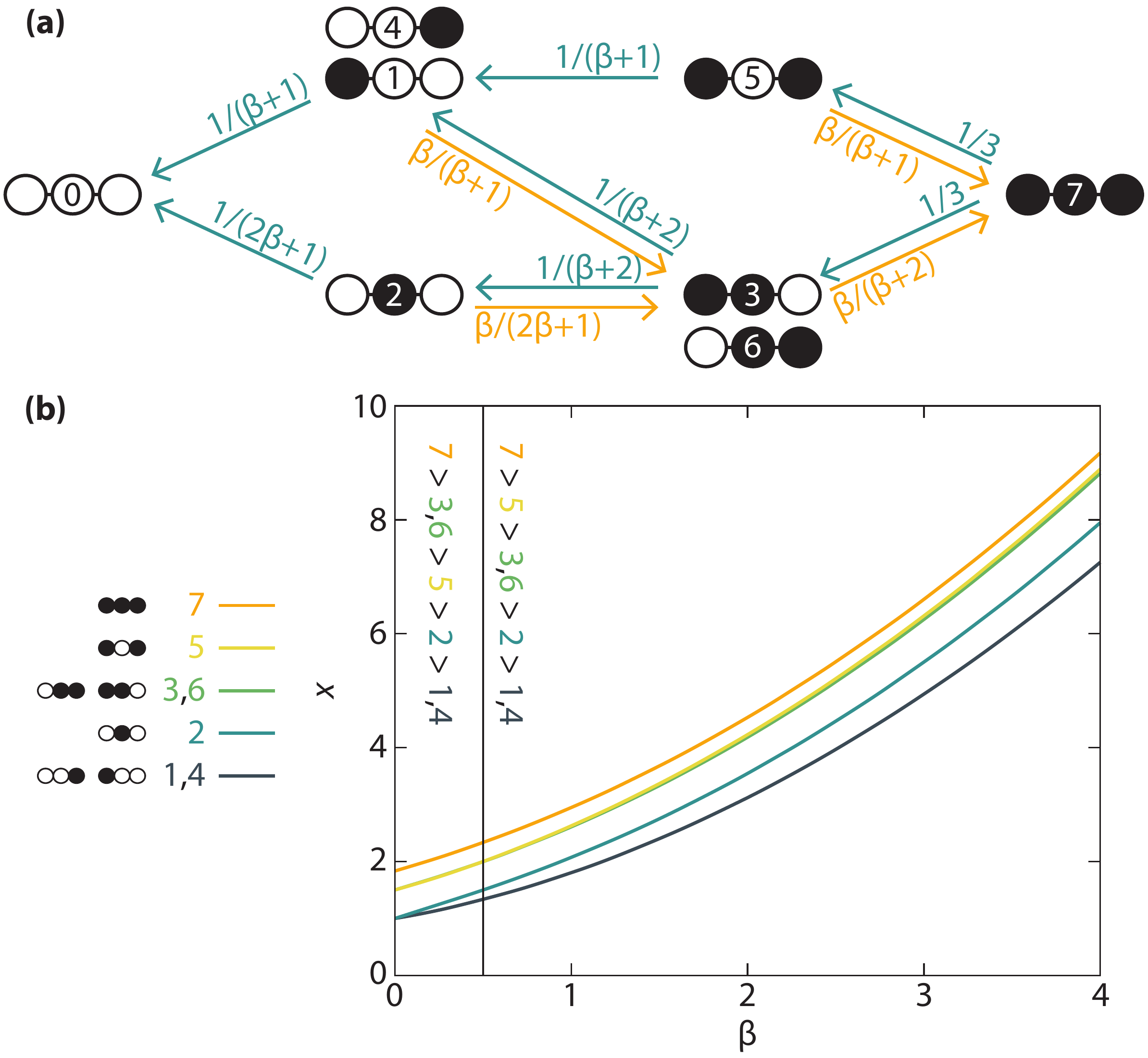}
\caption{(Color online) Panel (a) shows the four equivalence classes of configurations of the SIS model at the unique graph of three vertices and two edges. The values on arrows gives the transition probabilities. Arrows and probabilities for equivalent configurations (1 and 4, and 3 and 6) are only shown for one of the configurations. Configuration 0 is absorbing---no arrows lead out from it. Panel (b) shows the expected extinction times $x$ derived from (a) as a function of the infection rate $\beta$. The vertical line at $\beta=1/2$ shows where configuration 5 start having a longer expected extinction time than configurations 3 and 6.}
\label{fig:ex}
\end{figure}

\subsection{Kendall's $\tau$}

We compare several types of correlations (e.g.\ between structural measures and times to extinction) in this work. To do that, we will use Kendall's $\tau$, a rank-type correlation coefficient. It is defined as the fraction of pairs connected by a line with a positive slope, minus the fraction of pairs connected by a negative slope~\cite{knight}. If its value is $+1$, there is a perfect correlation between the ranks of all data points; if the value is $-1$, there is a perfect anti-correlation; $\tau=0$ represents no correlation. We use this coefficient rather than other popular ones for three reasons. First, the output data is typically not Gaussian, so the premises for Pearson's correlation coefficient is violated. Second, to reduce the disk space usage we do not store the explicit expressions of $\mathbf{x}$, but rather the order of them in the large and small $\beta$ limits (the actual values are not needed to calculate $\tau$, only the rank). Third, the number of data points is small enough to use Kendall's $\tau$ rather than the faster, but less principled, Spearman rank correlation coefficient.

\section{Results}

\subsection{An example}

We start the discussion of our results by examining the example of Section~\ref{sec:exptime} and Fig.~\ref{fig:ex}. Many properties of the solution, Eqs.~\eqref{eq:ex_res}, hold also for other $N$.

First, in the small limit of $\beta$, the solutions are the harmonic numbers of $\omega_s$. This follows immediately from the dynamics defined above---all events are recovery events, the time to the next event is $1/\omega_t$, where $\omega_t$ decreases by one every event, leading to the harmonic number $\sum_{t=1}^s 1/\omega_t$.

Second, for large $\beta$, the extinction time approaches the asymptote $u\beta^{N-1}$. For all graphs we study, $u$ is constant with respect to $s$ but dependent on graph structure $G$. Below, we study $u$ for all our graphs.

\subsection{Solving our example with Cramer's rule}

In this section, we introduce Cramer's rule as a way to solve the extinction times. This is a computationally inefficient method, but the way to get some analytic insights into the asymptotic behavior of $u\beta^{N-1}$ (as we will in subsequent sections). Let us derive this exponent for Eqs.~\eqref{eq:ex_res}. To do this, we apply Cramer's rule to the polynomial matrix $\mathbf{Y}(\beta)$ denoted further as $\mathbf{Y}$ (we will drop the $\beta$ argument for most of the derivation below). Cramer's rule states that the $s$'th element of vector $\mathbf{x}$ from Eq.~\eqref{eq:self2} is
\begin{equation}\label{eq:solx}
x_s  = \frac{\det \mathbf{Y}^s}{\det \mathbf{Y}},
\end{equation}
where $\mathbf{Y}^s$ is a matrix obtained from $\mathbf{Y}$ by replacing the $s$'th column by the vector $-\mathbf{1}$ (i.e.\  all elements being minus one).

Let us consider $x_{1,4}$ for our example above ($x_1$ and $x_{4}$ are identical since $1$ and $4$ are automorphic). We will use the row and column indices of ${\bf Y}$ in this section.

In order to calculate the polynomial degree of determinant of matrix $\mathbf{Y}^s$ we first make a subfactor expansion along the first column of matrix $\mathbf{Y}$. This gives the expressions:
\begin{subequations}\label{eq:yy1}
\begin{align}
\det {\bf Y} &= (-\beta-1) M_{11} + M_{31} - 2M_{41}\\\det {\bf Y}^1 &= - M_{11} + M_{21} - M_{31}+ M_{41}- M_{51}
\end{align}
\end{subequations}
where $M_{st}$ is the determinant of the matrix {\bf Y} without $s$'th row and $t$'th column (i.e.\ the  $st$\textit{-minor} of {\bf Y}). We find that
\begin{subequations}
\begin{align}
M_{11} &=12 \beta^2 + 22 \beta + 12 \\
M_{21} &= - 4 \beta^2-6\beta \\
M_{31} &= 8 \beta^3 + 16 \beta^2 +6\beta   \\
M_{41} &=- 2 \beta^3- \beta^2  \\
M_{51} &= 4 \beta^4 + 6 \beta^3 + 2 \beta^2  ,
\end{align}
\end{subequations}
giving (via Eq.~\eqref{eq:yy1}):
\begin{subequations}
\begin{align}
\det {\bf Y} &= - 16 \beta^2 - 28 \beta - 12 \\ \det {\bf Y}^1 &= -4\beta^4-16\beta^3- 35 \beta^2 - 34 \beta-12 ,\label{eq:subf}
\end{align}
\end{subequations}
which is in agreement with the numerical results from Eq.~\eqref{eq:ex_res}.

\subsection{Asymptotic scaling: exact relations}

We will prove that the leading term of $x_s$ is $u\beta^c$ for an integer $c$. For all our 12,110 graphs and all 2,963,056 configurations, we have  $c=N-1$. We believe this holds in general, but we have to leave a proof of that for the future.

From Eq.~\eqref{eq:solx}, we see that our assertion will be true if we can show that the leading term of $\det {\bf Y}^s$ is independent of $s$. In the Appendix, we show that the determinants of the $ns$-minors of ${\bf Y}$ (cf.\ Eq.~\eqref{eq:subf}) have leading terms of polynomial degree $n-1$ and the same prefactor, independent of $s$.  Such a large polynomial degree is impossible to attain for $st$-minors with $s<n$, since they have rows not containing any $\beta$, some of the $n-1$ factors of the Leibniz expansion of the determinant must have polynomial degree zero with respect to $\beta$. Thus the leading behavior of $\det {\bf Y}^s$ comes from $M_{ns}$ and is unique. Since $\det {\bf Y}$ is trivially independent of $s$, the leading behavior of $x_s$ is also $s$-independent. If $c=N-1$, we can conclude that $\det{\bf Y}=n-N+1$

For our example graph (and some other simple graphs of $N\leq 4$ we check) it holds that 
\begin{equation} \label{eq:hypo}
\deg(M_{st}) = n - N + \omega_s ,
\end{equation}
for all states $t$. If this is true in general, then, curiously, $\det {\bf Y}$ is determined by the minors corresponding to the configurations with lowest prevalence and $\det {\bf Y}^s$ the one with the highest prevalence. This is reminiscent of current-flow networks where the determinant of the $st$-minor of the adjacency matrix is proportional to the potential drop between $s$ and $t$ \cite{detcurrflow}. 

\begin{figure}
\includegraphics[width=\columnwidth]{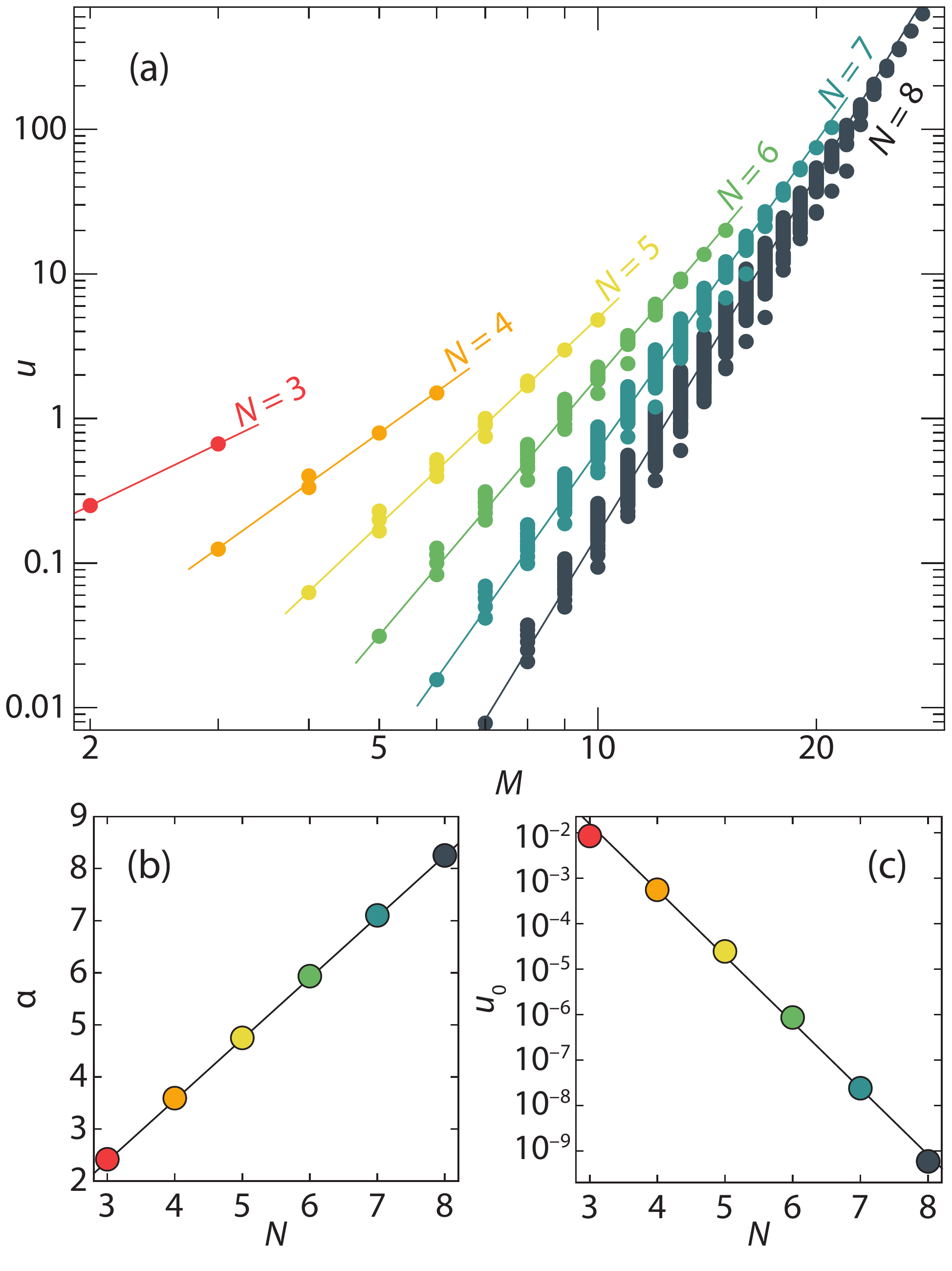}
\caption{(Color online) The size-dependence of the asymptote $u$ as a function of the number of edges $M$. In panel (a), we see a power-law dependence of $u=u_0 M^\alpha$ on the number of edges given the number of vertices. In panels (b) and (c), we see the $N$ dependence of parameters $u_0$ and $\alpha$.}
\label{fig:leading_vs_m}
\end{figure}

\begin{figure}
\includegraphics[width=0.7\columnwidth]{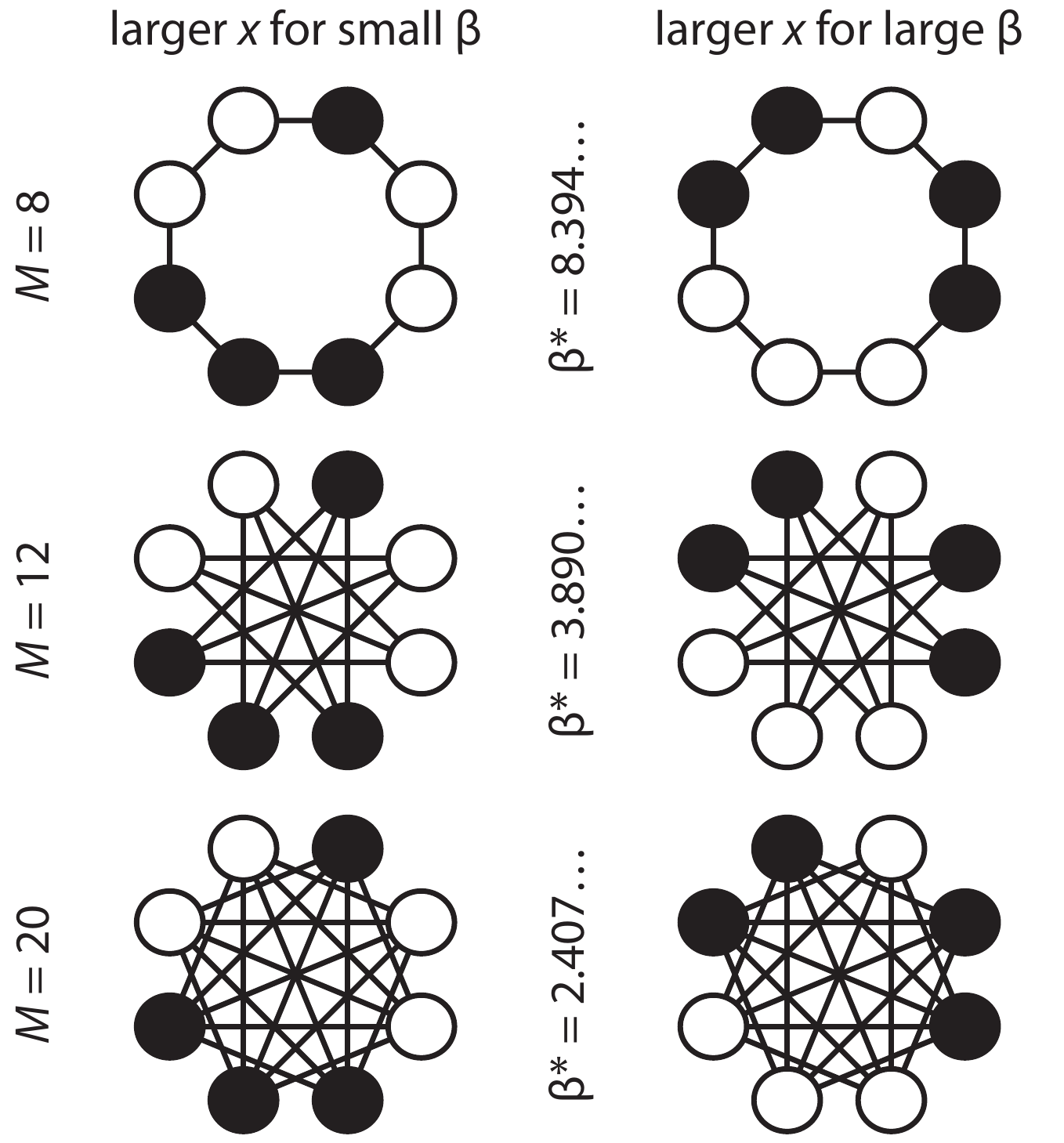}
\caption{The only three graphs in our study where all vertices are in equivalent positions but the ranking of configurations (in order of extinction time) depends on $\beta$. Black represents infectious; white represents susceptible. $\beta^\ast$ gives the $\beta$ value where the two configurations have the same expected extinction time.
}
\label{fig:diff}
\end{figure}

\begin{figure*}
\includegraphics[width=\textwidth]{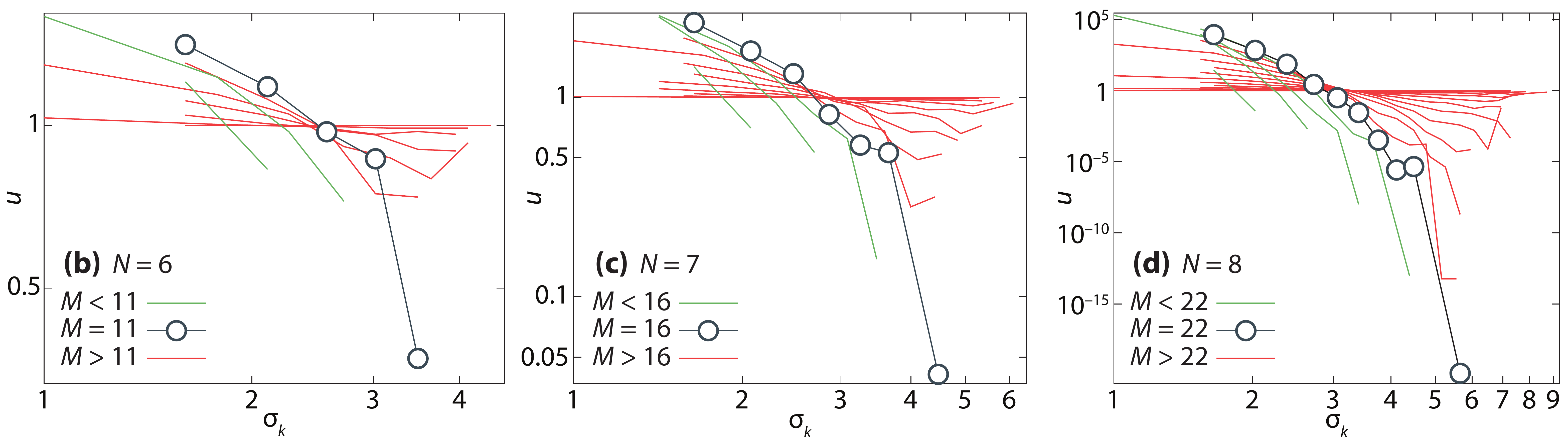}
\caption{(Color online) The asymptotic coefficient $u$ as a function of the standard deviation of the degree of the vertices. Different panels represent different number of vertices ($N\geq 6$); different curves represent different number of edges. The curves with the smallest $u$-values, given $N$, are highlighted ($M=11,16,22$) as a reference. Note that the axes are logarithmic.}
\label{fig:vssd}
\end{figure*}

\subsection{Numerical results for the large $\beta$ asymptotics}
\label{sec_larg_b}

As mentioned above, the $\beta\rightarrow\infty$ behavior is the same for all configurations $s$, namely $x_s= u\beta^{N-1}+O(\beta)$. In this section, we investigate how the sizes of the graphs control the prefactor $u$.

As we can see in Fig.~\ref{fig:leading_vs_m}(a), for a given $N$, $u$ is a power-law of the number of edges $M$ (keeping in mind that $u$ depends on the graph structure):
\begin{equation}
u(M)=u_0 M^\alpha~.
\end{equation}
For the graphs we study, the coefficients $\alpha$ and $\ln u_0$ have a close to linear dependence of $N$ (Fig.~\ref{fig:leading_vs_m}(b) and (c)):
\begin{subequations}\label{eq:u0_alpha}
\begin{align}
u_0 & = & 126(1) \times 0.0268(2)^{N-1}   \\
\alpha & = & -1.081(2) + 1.168(1) N
\end{align}
\end{subequations}
where the number in parentheses represents the standard errors in the last digit. Of course, the error estimates are based on the data we have and subjected to small-size effects. In other words it is conceivable that $\alpha$ could be taken as $N-1$, giving a large-$\beta$ approximation $\hat{x}$ of the extinction time $x$:
\begin{equation}
\hat{x}(\beta,N,M) = a(b\beta M)^{N-1}
\end{equation}
with constants $a\approx 126$ and $b\approx 0.0268$. Note also that there is a weak but consistent bend (negative second derivative) of $\ln u_0$ as a function of $N$ (i.e.\ $a$ and $b$ seem to be slowly varying functions of $N$).

\begin{table}
\caption{Correlation between measures characterizing the structure of graphs (beyond the number of vertices and edges) and the large-$\beta$.\label{tab:corr}}
\begin{ruledtabular}
\begin{tabular}{ll}
Measure & \text{Kendall's~} $\tau$ \\ \hline
Clustering coefficient & $-0.667$ \\
Degree assortativity & $0.191$ \\
Average distance & $-0.309$ \\
S.d.\ of degrees & $-0.751$ \\
\end{tabular}
\end{ruledtabular}
\end{table}

As seen in Fig.~\ref{fig:leading_vs_m}, $u(G)$ is not completely determined by Eq.~\eqref{eq:u0_alpha}---there is also some spread of the points for a given $N$ and $M$. To understand what causes two graphs of the same $N$ and $M$ to differ, we try several structural predictors: the clustering coefficient (a.k.a.\ transitivity---the fraction of triangles among all connected subsets of three vertices), the degree assortativity (the Pearson correlation of degrees at either side of an edge), the average distance ($d(i,j)$---the fewest number of edges of any path between $i$ and $j$), and the standard deviation of the degree. See Refs.~\cite{mejn:book,barabasi:book} for detailed descriptions of our measures. For all pairs of $N$ and $M$, we calculate the correlation between the $u$ and these measures, then we average these values over all graphs. The results, shown in Table~\ref{tab:corr} shows that all correlations, except the one with degree assortativity, are negative and the strongest correlation is with the standard deviation of degree $\sigma_k$. This means epidemics in graphs with more homogeneous degree sequences tend to last longer in the large $\beta$ limit. The relationship between $u$ and $\sigma_k$ is shown explicitly in Fig.~\ref{fig:vssd}. Indeed, for every combination of $N$ and $M$ the $u$ vs.\ $\sigma_k$ curves are almost always decaying. We highlight the curves with largest range in $\ln u$, and note that these occur for close to maximally dense graphs.

\begin{figure}
\includegraphics[width=0.8\columnwidth]{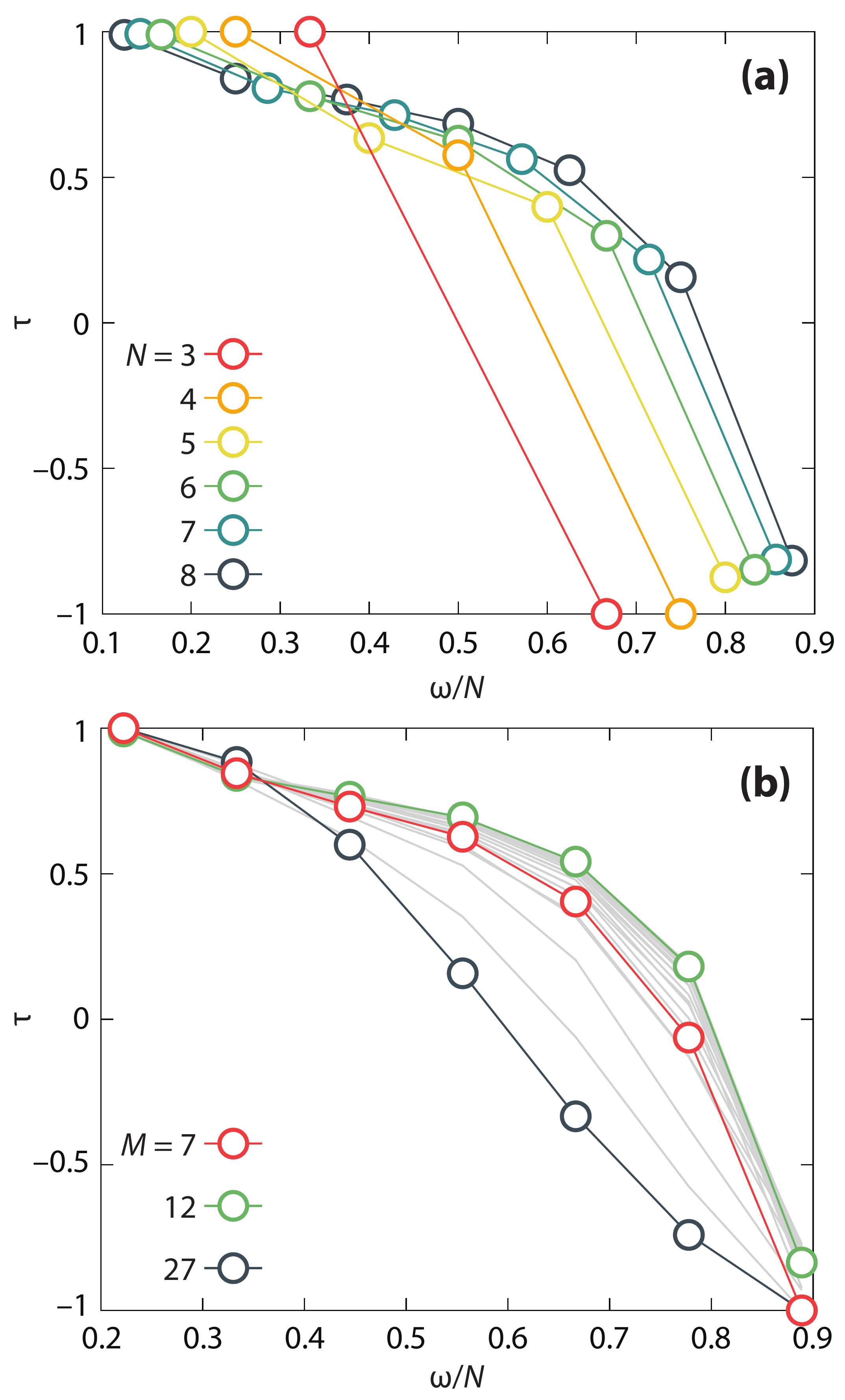}
\caption{(Color online) The average correlation (Kendall's) $\tau$ of the ranking of configurations in the high and low limits of $\beta$ as a function of the relative prevalence $\omega_s/N$. Panel (a) shows values averaged over all connected graphs of a certain number of vertices; panel (b) shows the same quantity for $N=8$ with curves representing averages over graphs with a certain number of edges. The gray curves represent the $M$ values not discussed in the text.}
\label{fig:dev}
\end{figure}

\begin{figure}
\includegraphics[width=0.8\columnwidth]{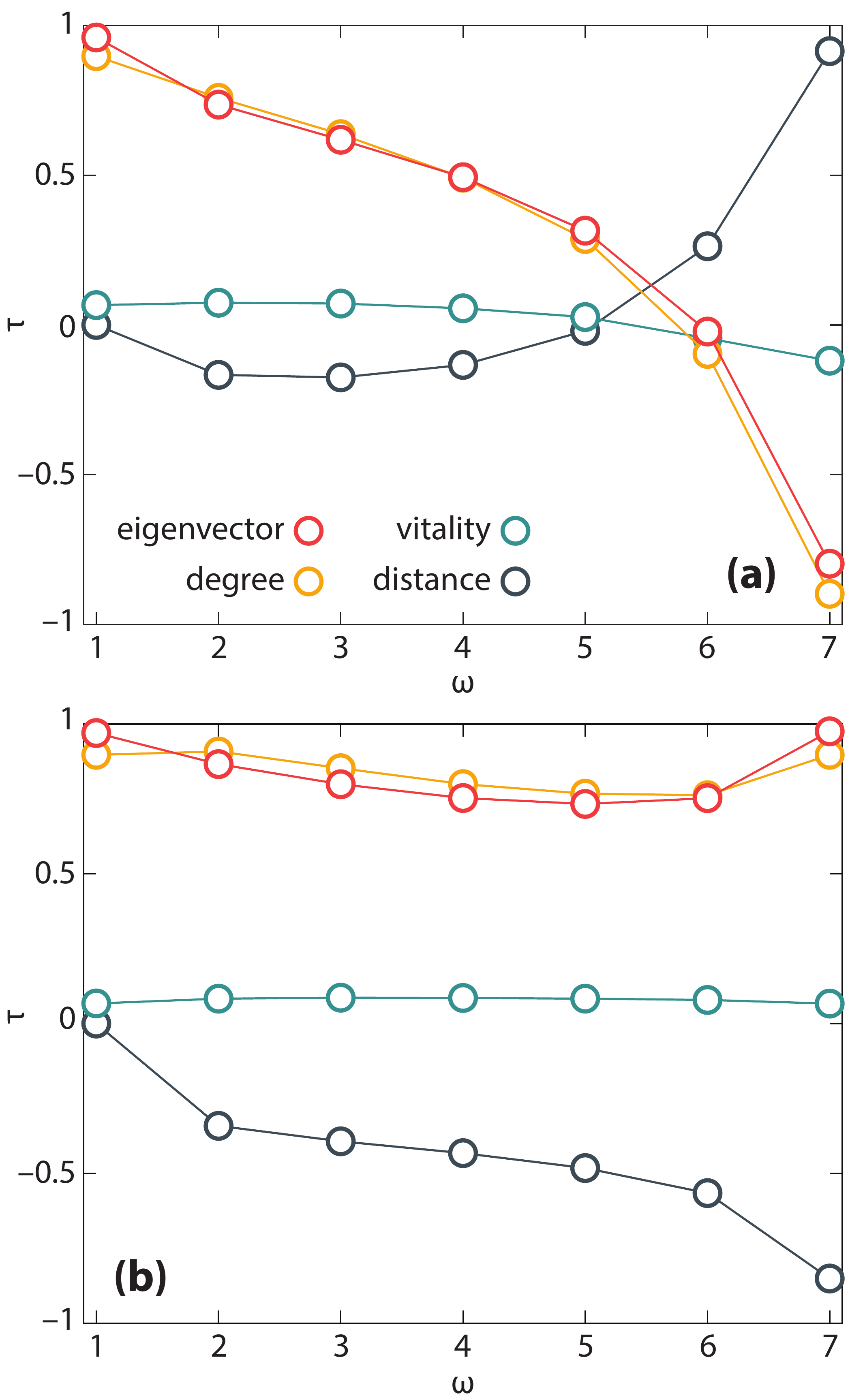}
\caption{(Color online) The average correlation (Kendall's $\tau$) between the rank of the configurations at large (a) and small (b) $\beta$ and various measures of the position of infectious vertices as function of the number $\omega$ of infectious vertices. These curves are averaged over all connected graphs with $N=8$ and $M=14$ (i.e.\ with a connectance---fraction of vertex pairs being an edge---$1/2$).}
\label{fig:ycorr}
\end{figure}

\subsection{Pervasiveness of $\beta$-dependent rankings of configurations}

Already from Fig.~\ref{fig:ex}, we know that the ranking of configurations with the same number of infectious vertices can depend on $\beta$. As it turns out, for all but 20 of the 12,110 graphs we study, there is at least one pair of configurations, where one has a longer expected time to extinction for small $\beta$ and the other for large $\beta$. The exceptions to this are all graphs where all vertices are automorphically equivalent. There are 23 such graphs among the ones we study. The three exceptions to the exceptions---the only configurations among such symmetric graphs with a $\beta$-dependent ranking---are shown in Fig.~\ref{fig:diff}. There are some interesting symmetries between these graphs evident from the figure. For example, even though the $M=8$ and $M=20$ graphs are complements to each other---two pairs of vertices in one case has an edge if and only if it does not in the other---the ranking of the configurations (which one is dominant for large vs.\ small $\beta$) is the same. However, we do not have any explanations for this observation. The actual times to extinction are extremely similar between the two configurations. In the $M=8$ case, for example, the numerator of $x$ for the left configuration starts as:
\begin{eqnarray}
 97844723712\times\beta^{28}&+
& 2019406381056\times\beta^{27}+\nonumber \\
 20485144313856\times\beta^{26}&+
& 136322491613184\times\beta^{25}+\nonumber\\ 
 6704\mathbf{61968908288}\times\beta^{24}&+&\dots
\end{eqnarray}
while the right configuration has the numerator:
\begin{eqnarray}
 97844723712\times\beta^{28}&+
& 2019406381056\times\beta^{27}+\nonumber \\
 20485144313856\times\beta^{26}&+
& 136322491613184\times\beta^{25}+\nonumber\\ 
 6704\mathbf{55853613056}\times\beta^{24}&+&\dots
\end{eqnarray}
 (with the differences highlighted by bold face). Needless to say (since they also share the small-$\beta$ asymptotics) plotting them in the same graph does not show any visible difference. This is an example of a result that would be almost impossible to detect by stochastic simulations.

As $N$ increases, there are more opportunities for symmetry breaking configurations with the same $\omega_s$ (below, where it is clear from the context, we write simply $\omega$). One scenario is that for even larger $N$, the only graphs with $\beta$-independent rankings are the fully connected graphs (because for them, all configurations of the same $\omega$ are automorphically equivalent and this hence simplifies the system of equations from Section \ref{sec:exptime}). We note that fully connected graphs are the most common interaction structures studied in the literature, and perhaps an unfortunately atypical case.

\subsection{Correlation of asymptotic behaviors}

In this section, we continue the investigation of the $\beta$-dependence of the rankings of expected extinction times. After calculating $\mathbf{x}$, we rank the configurations in the limits of large and small $\beta$. Let $r_L(s,G)$ be the normalized rank of configuration $s$ among all configurations of the same prevalence $I$ in the large $\beta$ limit; and $r_S(s,G)$ the corresponding quantity as $\beta\rightarrow 0$. Then we use Kendall's $\tau$ coefficient to measure the correlation between $r_L$ and $r_S$. 

In practice, we first put all $x_s$ on the minimal common denominator and compare the numerators (which are polynomials with integer coefficients). To rank polynomials in the small $\beta$ limit, one first compare the constant coefficient (which is the same for all configurations of the same prevalence), then we use coefficients of increasing polynomial degree as tie-breakers. To rank polynomials as $\beta\rightarrow\infty$, one goes through the coefficients in the opposite direction---one polynomial is considered larger than the other if the highest order coefficient where they differ is larger. As the full equations of the solution for $x_s(\beta)$ take much disk space, we only save the rankings.

With the ranking of the solutions at hand, we proceed to calculate $\tau$ (using the NumPy library of Python). In Fig.~\ref{fig:dev}(a), we see $\tau$ as a function of the prevalence $\omega$ averaged over all connected graphs of given number of vertices $N=3,\dots,8$. $\tau$ is strictly decreasing from $+1$ to $-1$. The decrease (in particular for larger $N$) is faster in the beginning and end than in the middle. Since the curve gets consistently less steep with $N$ for intermediate prevalence values, it seems possible that the curve would flatten out with a growing $N$. In other words, for configurations with few infectious vertices, the ranking is rather independent of $\beta$; while for high-prevalence configurations, all curves of expected time to extinction will cross as $\beta$ increases.

In Fig.~\ref{fig:dev}(b), we take a closer look at the $N=8$ graphs and split the average into different curves depending on the number of edges. $\tau$ has an intermediate maximum for $M=12$ while the sparsest graph ($M=7$) has smaller $\tau$ than the densest ($M=27$). For graphs close to the maximum number of edges, the region of slower decrease for small $\omega$ is almost gone.

\subsection{Structural determinants of the asymptotic behavior}
\label{sec_struc_det}

From the analysis of Fig.~\ref{fig:dev}, we know that the number of vertices and edges affect the ranking of configurations. Now we will look at more detailed  explanations based on graph structure---what determines the ranking for graphs of the same $N$ and $M$? Fig.~\ref{fig:ycorr} presents a case study for $N=8$ and $M=14$ (when exactly half of the vertex pairs are connected by an edge).

To measure the correlation, we once again use Kendall's $\tau$. We pick four structural measures to characterize a configuration. Then we correlate each one with either $r_S$ or $r_L$. We present these measures briefly below. For a thorough account, we refer to Refs.~\cite{barabasi:book,mejn:book}.
\begin{enumerate}
\item Average \textit{degree}---number of neighbors---of the infectious vertices. Degree is the simplest notion of centrality, but also local (in the sense that a vertex' degree is only dependent on its neighborhood).
\item Average \textit{eigenvector centrality} over the infectious vertices. The eigenvector centrality is given by the eigenvector corresponding to the leading eigenvalue of the adjacency matrix. It is perhaps the most straightforward generalization of degree to account for the idea that being close to central vertices makes a vertex central.
\item Average \textit{vitality} of the infectious vertices. Vitality is the general class of measures based on measuring the response of some graph descriptor on the deletion of a vertex~\cite{koschutzki2005centrality}. Following Ref.~\cite{holme3f}, we define the vitality (technically \textit{component-size vitality}) of vertex $i$ to be $v(i) = [S(G)-1]/S(G\setminus \{i\})$. Where $S(G)$ denotes the number of vertices of the largest connected component of graph $G$. This measure will be very close to one for larger graphs and would thus be unsuitable if one were to scale this study up. It is, on the other hand, interesting as it is directly measuring the contribution the presence of a vertex makes in a worst case, $\beta\rightarrow\infty$, scenario. 
\item Average \textit{distance} $d(i,j)$ between infectious vertex pairs $i\neq j$. This is the only measure we use that does not involve averaging a centrality measure.
\end{enumerate}
In addition to these we also try the standard measures \textit{betweenness} and \textit{closeness} centrality, but these do not contribute much to the understanding. Partly because they are very correlated with vitality and eigenvector centrality, respectively, for these small graphs. Partly because their rationales involves flows along shortest paths between random pairs---a type of process not present in disease spreading.

In Fig.~\ref{fig:ycorr}, we plot results of the correlation coefficient between the the above measures of position and the ranks of the configurations for all connected graphs with $N=8$ and $M=14$. Fig.~\ref{fig:ycorr}(a) shows the case of large $\beta$. For configurations with low prevalence, we see that large degree is strongly correlated with the extinction-time ranking. This is natural, since for low $\beta$ and $\omega$ secondary effects are negligible---the number of ways to increase $\omega$ counts more than other factors, i.e.\ the degree. Eigenvector centrality behaves almost like degree. Although not identical, these two quantities are strongly correlated for the small graphs we study, so it is natural the values are close. Somewhat interestingly, which one of these that gives the largest $\tau$ varies with $\beta$ in an irregular way. For sparser graphs (than the one in Fig.~\ref{fig:ycorr}), that are more prone to disintegrating upon vertex-deletion, vitality shows a correlation on par with degree and eigenvector centrality. Finally, the average distance shows an increasing correlation with $\omega$. That the infectious vertices are far away means that the surface to the susceptible vertices are larger and thus that the next event is more likely to be an infection event.

The structural correlations for the small-$\beta$ case (Fig.~\ref{fig:ycorr}(b)) is no surprise in the light of Fig.~\ref{fig:dev} and Fig.~\ref{fig:ycorr}(a). Since there is a correlation between $r_S$ and $r_L$ at small $\omega$ and a corresponding anti-correlation at large $\omega$, we expect the correlations with structural measures to be similar between the small and large $\beta$ cases for small $\omega$ and different for large $\omega$. This is also rather accurately describing what happens. In this case, the degree and eigenvector centrality are strongly correlated with $r_S$ for all $\omega$, while the distance becomes strongly anti-correlated for large $\omega$.

\section{Discussion}

We have studied the extinction of SIS epidemics on small graphs. We did so by calculating the exact expressions for the expected time to extinction for all connected graphs between three and eight vertices.

We find that, for a given graph, the limit behavior as $\beta\rightarrow\infty$ is independent of the configuration of susceptible and infectious, while for $\beta\rightarrow 0$ it is (trivially) just dependent on the number of infectious vertices. Of course both these limits are not of an immediate practical interest, but to understand the in-between reality we need to understand the extremes.

The large-$\beta$ asymptote $u$ of the extinction time depends on the size and structure of the graph---for large $\beta$ we find $u = u_0M^\alpha\beta^{N-1}$ where both $u_0$ and $\alpha$ are linear functions of $N$ for the graphs we investigate. Our final formula for the large-$\beta$ behavior of $x$ is $a(b\beta M)^{N-1}$ where $a\approx 126$ and $b\approx 0.0268$. This super-exponential $N$-dependence is in line with earlier observations~\cite{nasell,doering,o2,shadows,fagani,schwartz}. Simply speaking, even though there is a finite chance of extinction of SIS epidemics in finite graphs, for $\beta$ only a little more than one, this probability is so small it can be ignored for all practical purposes for all but the smallest graphs. For graphs of the same number of vertices and edges, the strongest determinant of the asymptotic behavior of the time to extinction is the variability (measured by the standard deviation) of the degree. Finally, we find that outbreaks tend to last longer in graphs of heterogeneous degree distributions. 

Furthermore, given an interaction graph, we investigated when the ranking of configurations with respect to extinction time changes. For configurations with few infectious vertices, the rankings are typically the same for large and small $\beta$; for configurations with many infectious vertices, there is an anti-correlation between the extinction times in the large and small $\beta$ limits. The main structural predictors for the rank of configurations with the same number of infectious vertices (for the same graph) are degree and eigenvector centrality, while the correlations with vitality and inter-vertex distance are weaker.

The main contribution of this paper is to give a view of the relation between graph structure and epidemic behavior from another perspective than the usual. Rather than studying the $N\rightarrow\infty$ limit by stochastic simulations, we study exact expectation values of small graphs. This enables us to discover hypotheses that could be tested in standard stochastic simulations. It also makes it possible to discover the smallest graphs and configurations with some specific properties. For example, the graph of Fig.~\ref{fig:ex}(a)---where the configurations 3 and 6 have longer extinction times than configuration 5 in the interval $0<\beta < 1/2$ and vice versa for $\beta > 1/2$---is the smallest of a graph where configurations change the order of expected extinction time with $\beta$. The answer to the reversed question---what is the smallest graph where all same-prevalence configurations are ranked equally for large and small $\beta$---is the triangle $E=\{(0,1),(1,2),(2,0)\}$. In fact, all graphs where at least two vertices are in different positions (not automorphically equivalent) do not have the equal-ranking property, but 20 of 23 such graphs we study where all vertices are equivalent do have it. On one hand, these observations do not generalize, and thus follow more of a mathematical mode of scientific exploration. On the other hand, they are the basis of hypotheses that could be tested by future theory. For example, for every graph where not all nodes are equivalent, will the ranking of configurations always depend on $\beta$?

We anticipate more computational epidemiology studies without random numbers in the future, and simulation studies testing the findings in this work holds for larger graphs. It would also be interesting to go beyond expected times and derive the probability distribution of extinction time. That would need a different computational approach. For a model of networks one could consider mapping the problem to that of mean first-passage times~\cite{iannelli,fpt,masuda:rev}, or combinatorial stochastic processes~\cite{raaz}.

\begin{acknowledgments}
We thank Naoki Masuda and Petteri Kaski for helpful comments. PH was supported by JSPS KAKENHI Grant Number JP 18H01655. LT acknowledges the support under Grant No.\ ANR--13--JSV5--0006--01 of the French National Research Agency.
\end{acknowledgments}

\appendix

\section{Leading terms of minors are identical} 

In this appendix, we will prove that the leading terms of the $ns$-minors of $\mathbf{Y}$ are the same. For this proof, first recall Leibniz formula for determinants saying that a determinant is a sum of products of matrix elements
\begin{equation}\label{eq:leibniz}
\prod_{s=1}^n{Y}_{s\sigma(s)}
\end{equation}
where $\sigma(s)$ is a permutation of the numbers to $n$ (the size of ${\mathbf{Y}}$, i.e.\ the number of automorphic equivalence classes of configurations). To calculate the full determinant one needs to multiply half of the terms by $-1$, but we can ignore that for our purpose. We will show that the $ns$-minors have exactly one term with polynomial degree $n$ equal to the product of the prefactors of $\beta$ along the diagonal of $\mathbf{Y}$.

Recalling points \ref{item:diag} and \ref{item:row} of Section~\ref{sec:exptime}, we can write $Y_{ss}=-B_s\beta-A_s$ with $A_s,B_s>0$ for $s=1,\dots,n-1$. Furthermore, there is a number $j(s)>s$ such that $Y_{sj(s)}=B_s$. For the special case $M_{nn}$ our results is easy since the $nn$-minor is upper triangular with respect to elements containing $\beta$, and all diagonal elements contain $\beta$. Clearly the the leading coefficient is the product of the diagonal, $\prod_{s=1}^{n-1}B_s$, since any other term would contain sub-diagonal elements with polynomial degree zero.

We will solve the case $s<n$ by constructing an algorithm to find the unique leading term of $M_{ns}$. We will work with the indices of $\mathbf{Y}$ rather than indices of the minor.
\begin{enumerate}
\item Set $\Lambda:=\{s\}$, $i:=j(s)$ and $z:=1$.
\item \label{item:if} If $i=n$, go to step \ref{item:exit}.
\item \label{item:multiply} Multiply ${Y}_{ij(i)}$ to $z$, add $i$ to $\Lambda$. 
\item \label{item:update} Set $i:=j(i)$ and go to step \ref{item:if}.
\item \label{item:exit} For all $i\in\Lambda\cap \{1,\dots,n\}$, multiply ${Y}_{ii}$ to $z$.
\end{enumerate}
$z$, at the exit, is a the term of $M_{ns}$ with highest polynomial degree. First, we note that by construction, $z$ is a product of $n$ elements of the $ns$-minor of $\mathbf{Y}$. Then, because the row and column indices are strictly increasing when updated at step~\ref{item:update}, no row or column index of an element multiplied by $z$ at step~\ref{item:multiply} occur twice. Furthermore, this is also true at step~\ref{item:exit} (otherwise they such elements would already be multiplied into $z$ are step~\ref{item:multiply}), so $z$ is indeed a term of $M_{ns}$. It has polynomial degree $n-1$, which is the maximal possible since the maximal polynomial degree of matrix $\mathbf{Y}$ elements is one. Moreover, there cannot be any other term of $M_{ns}$ with polynomial degree $n-1$. Step \ref{item:multiply} adds factors that must necessarily belong to a term of polynomial degree $n-1$ (since there is only one element containing $\beta$ at row $s$). Finally, one cannot multiply by an element ${Y}_{ij(i)}$ rather than ${Y}_{ii}$ at step~\ref{item:exit}, since then $z$ would not include an element from the $i$'th column.

Unfortunately, it is not straightforward to extend this algorithm to a proof of Eq.~\eqref{eq:hypo}. For example, for $st$-minors with $t<n$, there can be columns without any element containing $\beta$ meaning that the leading terms (that then will have a polynomial degree less than $n-1$) can contain elements from below of the diagonal $\mathbf{Y}$. This means that one cannot base proofs about the algorithm on the fact that it samples elements of increasing indices as above.

\bibliographystyle{abbrv}
\bibliography{exact}

\begin{thebibliography}{10}

\bibitem{andersonmay}
R.~M. Anderson and R.~M. May.
\newblock {\em Infectious diseases in humans}.
\newblock Oxford University Press, Oxford, 1992.

\bibitem{Bajardirsif20120289}
P.~Bajardi, A.~Barrat, L.~Savini, and V.~Colizza.
\newblock Optimizing surveillance for livestock disease spreading through
  animal movements.
\newblock {\em J. Roy. Soc. Interface}, 9(76):2814--2825, 2012.

\bibitem{barabasi:book}
A.-L. Barab\'asi.
\newblock {\em Network Science}.
\newblock Cambridge Press, Cambridge UK, 2016.

\bibitem{detcurrflow}
R.~L. Brooks, C.~A.~B. Smith, A.~H. Stone, and W.~T. Tutte.
\newblock Determinants and current flows in electric networks.
\newblock {\em Discrete Mathematics}, 100:291--301, 1992.

\bibitem{vf2}
L.~P. Cordella, P.~Foggia, C.~Sansone, and M.~Vento.
\newblock An improved algorithm for matching large graphs.
\newblock In {\em 3rd IAPR-TC15 Workshop on Graph-based Representations in
  Pattern Recognition, Cuen}, pages 149--159, 2001.

\bibitem{igraph}
G.~Cs\'ardi and T.~Nepusz.
\newblock The igraph software package for complex network research.
\newblock {\em InterJournal Complex Systems}, 1695:1695, 2006.

\bibitem{daleygani}
D.~J. Daley and J.~Gani.
\newblock {\em Epidemic modelling: {A}n introduction}.
\newblock Cambridge University Press, Cambridge, 1999.

\bibitem{doering}
C.~R. Doering, K.~V. Sargsyan, L.~M. Sander, and E.~Vanden-Eijnden.
\newblock Asymptotics of rare events in birth–death processes bypassing the
  exact solutions.
\newblock {\em Journal of Physics: Condensed Matter}, 19(6):065145, 2007.

\bibitem{fagani}
F.~Fagnani and L.~Zino.
\newblock Time to extinction for the sis epidemic model: new bounds on the tail
  probabilities.
\newblock {\em IEEE Transactions on Network Science and Engineering}, page~1,
  2017.

\bibitem{Gantmacher}
F.~R. Gantmacher.
\newblock {\em The theory of matrices}.
\newblock Chelsea Pub.\ Co., New York, 1960.

\bibitem{flint}
W.~B. Hart.
\newblock Fast library for number theory: {A}n introduction.
\newblock In {\em Proceedings of the Third International Congress on
  Mathematical Software}, ICMS'10, pages 88--91, Berlin, Heidelberg, 2010.
  Springer-Verlag.

\bibitem{hethcote}
H.~W. Hethcote.
\newblock The mathematics of infectious diseases.
\newblock {\em SIAM Rev.}, 42(4):599--653, 2000.

\bibitem{hindes}
J.~Hindes and I.~B. Schwartz.
\newblock Epidemic extinction paths in complex networks.
\newblock {\em Phys. Rev. E}, 95:052317, May 2017.

\bibitem{shadows}
P.~Holme.
\newblock Shadows of the susceptible-infectious-susceptible immortality
  transition in small networks.
\newblock {\em Phys. Rev. E}, 92:012804, Jul 2015.

\bibitem{holme3f}
P.~Holme.
\newblock Three faces of node importance in network epidemiology: Exact results
  for small graphs.
\newblock {\em Phys. Rev. E}, 96:062305, 2017.

\bibitem{fpt}
S.~Hwang, D.-S. Lee, and B.~Kahng.
\newblock First passage time for random walks in heterogeneous networks.
\newblock {\em Phys. Rev. Lett.}, 109:088701, Aug 2012.

\bibitem{iannelli}
F.~Iannelli, A.~Koher, D.~Brockmann, P.~H\"ovel, and I.~M. Sokolov.
\newblock Effective distances for epidemics spreading on complex networks.
\newblock {\em Phys. Rev. E}, 95:012313, Jan 2017.

\bibitem{heetae}
H.~Kim, S.~H. Lee, and P.~Holme.
\newblock Building blocks of the basin stability of power grids.
\newblock {\em Phys. Rev. E}, 93:062318, Jun 2016.

\bibitem{kiss}
I.~Z. Kiss, J.~C. Miller, and P.~L. Simon.
\newblock {\em Mathematics of Epidemics on Networks}.
\newblock Springer, Heidelberg, Berlin, 2017.

\bibitem{knight}
W.~R. Knight.
\newblock A computer method for calculating {K}endall's tau with ungrouped
  data.
\newblock {\em Journal of the American Statistical Association}, 61:436--439,
  1966.

\bibitem{koschutzki2005centrality}
D.~Kosch{\"u}tzki, K.~Lehmann, L.~Peeters, S.~Richter, D.~Tenfelde-Podehl, and
  O.~Zlotowski.
\newblock Centrality indices.
\newblock In {\em Network Analysis: Methodological Foundations}, pages 16--61.
  Springer, Berlin, Heidelberg, 2005.

\bibitem{masuda}
N.~Masuda.
\newblock Directionality of contact networks suppresses selection pressure in
  evolutionary dynamics.
\newblock {\em Journal of Theoretical Biology}, 258(2):323--334, 2009.

\bibitem{masuda:rev}
N.~Masuda, M.~A. Porter, and R.~Lambiotte.
\newblock Random walks and diffusion on networks.
\newblock {\em Physics Reports}, 716-717:1 -- 58, 2017.

\bibitem{mcclellan}
M.~T. McClellan.
\newblock The exact solution of systems of linear equations with polynomial
  coefficients.
\newblock {\em J. ACM}, 20(4):563--588, Oct. 1973.

\bibitem{McKay201494}
B.~D. McKay and A.~Piperno.
\newblock Practical graph isomorphism {II}.
\newblock {\em Journal of Symbolic Computation}, 60:94--112, 2014.

\bibitem{nasell}
I.~N{\aa}sell.
\newblock Extinction and quasi-stationarity in the {V}erhulst logistic model.
\newblock {\em Journal of Theoretical Biology}, 211(1):11 -- 27, 2001.

\bibitem{mejn:book}
M.~E.~J. Newman.
\newblock {\em Networks: An Introduction}.
\newblock Oxford University Press, Oxford UK, 2010.

\bibitem{o2}
O.~Ovaskainen and B.~Meerson.
\newblock Stochastic models of population extinction.
\newblock {\em Trends in Ecology \& Evolution}, 25(11):643--652, 2010.

\bibitem{ps_rmp}
R.~Pastor-Satorras, C.~Castellano, P.~Van~Mieghem, and A.~Vespignani.
\newblock Epidemic processes in complex networks.
\newblock {\em Rev. Mod. Phys.}, 87:925--979, Aug 2015.

\bibitem{qu_etal}
B.~Qu, C.~Li, P.~{Van Mieghem}, and H.~Wang.
\newblock Ranking of nodal infection probability in
  susceptible-infected-susceptible epidemic.
\newblock {\em Scientific Reports}, 7:9233, 2017.

\bibitem{raaz}
R.~Sainudiin and D.~Welch.
\newblock The transmission process: A combinatorial stochastic process for the
  evolution of transmission trees over networks.
\newblock {\em Journal of Theoretical Biology}, 410:137 -- 170, 2016.

\bibitem{schwartz}
I.~B. Schwartz, L.~Billings, M.~Dykman, and A.~Landsman.
\newblock Predicting extinction rates in stochastic epidemic models.
\newblock {\em J. Stat. Mech.}, 2009(01):P01005, 2009.

\bibitem{Simon2011}
P.~L. Simon, M.~Taylor, and I.~Z. Kiss.
\newblock Exact epidemic models on graphs using graph-automorphism driven
  lumping.
\newblock {\em Journal of Mathematical Biology}, 62(4):479--508, Apr 2011.

\end{thebibliography}

\end{document}